\documentclass[a4paper,10pt]{article}
\usepackage[utf8]{inputenc}

\usepackage[max2]{authblk}

\usepackage{fullpage}
\usepackage{hyperref}

\usepackage{amsmath}
\usepackage{amssymb}

\usepackage{cite}

\usepackage[table]{xcolor} 

\usepackage{graphicx}
\usepackage{subfigure}

\title{\textbf{Testing nonlinear-QED at the future linear collider with an intense laser}}

\author[1]{Anthony Hartin\footnote{Speaker, \texttt{hartin@mail.desy.de}}}
\author[2]{Stefano Porto\footnote{\texttt{stefano.porto@desy.de}}}
\author[1,2]{Gudrid Moortgat-Pick\footnote{\texttt{gudrid.moortgat-pick@desy.de}}}

\affil[1]{\small DESY, Hamburg, Germany\normalsize}
\affil[2]{ \small II. Institut f\"{u}r Theoretische Physik, University of Hamburg, Germany\normalsize}

\begin{document}

\date{Talk presented at the International Workshop on Future Linear Colliders (LCWS13), Tokyo, Japan, 11-15 November 2013.}
\maketitle

\begin{abstract}
The future linear collider will collide dense $e^+e^-$ bunches at high energies up to 1 TeV, generating very intense electromagnetic fields at the interaction point (IP). These fields are strong enough to lead to nonlinear effects which affect all IP processes and which are described by strong field physics theory. In order to test this theory, we propose an experiment that will focus an intense laser on the LC electron beam post-IP. Similar experiments at SLAC E144 have investigated nonlinear Compton scattering, Breit-Wheeler pair production using an electron beam of 46.6 GeV. The higher beam energies available at the future LC would allow more precise studies of these phenomena. Mass-shift and spin-dependent effects could also be investigated.
\end{abstract}

\section{Introduction}

In recent times, there have been numerous studies of Particle Physics in intense electromagnetic fields and nonlinear QED. The concept of critical field $E_{\footnotesize\mbox{cr}\normalsize}=1.32\times10^{18}$ V/m at which the vacuum polarizes \cite{Sauter:1931zz,Heisenberg:1935qt,Schwinger:1951nm} drew the attention of theorists in studies of processes in intense fields since the beginning of Quantum Field Theory. These type of studies were further motivated by the subsequent birth of lasers, plasma physics, $e^+e^-$ colliders as well as astroparticle physics \cite{Kuznetsov:2004tb} and heavy ion collision experiments \cite{Greiner:1985ce}. \\

However, it is only in the last couple of decades that technological developments in collider and laser physics have allowed the possibility of testing some of these processes and effects. This has induced increasing attention to the subject of strong field physics \cite{Heinzl:2008wh,DiPiazza:2011tq}. Indeed, new or planned laser facilities like HiPER \cite{Hiper}, ELI \cite{ELI}, are going to reach laser intensities up to $10^{26}\,\mbox{W/cm}^2$, close to the  value of $10^{29}\,\mbox{W/cm}^2$ corresponding to $E_{\footnotesize\mbox{cr}\normalsize}$. The interpretation of the data of such experiments needs an accurate estimate of dissipative electromagnetic cascades undergone by high energy electrons, positrons and photons due to the interaction with multiple photons from the intense field of the laser \cite{Bulanov:2013cga}. \\

Electromagnetic fields approaching the $E_{\footnotesize\mbox{cr}\normalsize}$ could also be reached in the rest frame of colliding electrons and/or positrons at the interaction point (IP) of the future linear collider, generating beamstrahlung and other effects \cite{Yokoya:1991qz}, and should be taken into account in precision high energy physics \cite{Porto:2013pia}. \\

The vacuum instability by spontaneous electron-positron pair production and nonlinear effects in QED processes, which are expected to occur in the presence of an external electromagnetic field strength of the order  $E_{\footnotesize\mbox{cr}\normalsize}$, can be studied by colliding high energy electron and/or photons beams with an intense  laser \cite{McDonald:1986zz}. To that end, a SLAC experiment E144 at Stanford \cite{BulaE144Proposal,Bamber:1999zt} tested nonlinear QED by impinging a Terawatt laser on the 46.6 GeV electrons of the SLAC Final Focus Test Beam. Observed, were nonlinear processes in which multiple photons from the laser were absorbed. In particular, measured nonlinear Compton scattering ($e^-+n\omega\to e^-+ \gamma$) \cite{PhysRevLett.76.3116} and Breit-Wheeler pair production ($\gamma+n\omega\to e^{+}e^{-}$) \cite{PhysRevLett.79.1626} were in agreement with the theoretical predictions \cite{Nikishov:1964zza,Nikishov:1964zz,Narozhnyi1964,Nikishov1965}. \\

We propose to repeat and extend the SLAC-E144 experiment by exploiting the high energy electron and positron beams of the future linear collider by creating an intense laser particle beam interaction point in the extraction line. Strong field effects such as nonlinear Compton scattering and Breit-Wheeler pair production could be measured at higher precision. Additionally, the electron mass-shift and non-linear, spin-dependent effects could also be studied. \\

In this paper, after summarising the results of the processes studied at SLAC E144 in Section \ref{SecE144}, we describe the possibilities of a similar experiement implemented at the future linear collider in Section \ref{SecILC}. Mass shift-effects are described in Section \ref{SecMassShift} and spin-dependent effects in electron-laser interactions are reviewed in Section \ref{SectionSpin}. The paper ends with conclusions in Section \ref{SecConclusions}.

\section{Nonlinear processes at SLAC-E144}\label{SecE144}
A nonlinear QED process can be said to occur when multiple photons from an external field, such as that of a laser, contribute to the process. The following Lorentz-invariant dimensionless parameters are fundamental for the description of nonlinear QED processes (see for ex. \cite{RitusNauka,DiPiazza:2011tq}):
\begin{equation}
\eta=\frac{e\sqrt{(A_{\mu})^2}}{m},\,\,\,\,\,\,\,\,\,\,\,\,\,\,\,\,\,\,\Upsilon=\frac{e}{m^3}\sqrt{(F_{\mu\nu}p^{\nu})^2}=\eta\,\, \frac{p\cdot k}{m^2}\,,
  \end{equation}

where natural units $c=\hbar=1$ are used and where $A_{\mu},F_{\mu\nu}, k_{\mu}$ are the external electromagnetic field 4-potential, its field strength tensor and its momentum respectively. $p_{\mu}$ is the momentum of the particle propagating in the external electromagnetic field, $e$ and $m$ are the electron charge and its mass. \\

A value for the dimensionless nonlinearity parameter $\eta\gtrsim \mathcal{O}(1)$ indicates the onset of significant multiphoton effects. $\Upsilon$ is a nonlinearity parameter describing the magnitude of the field strength seen by an oncoming particle. A value of $\Upsilon\gtrsim \mathcal{O}(1)$ indicates that the field strength in the particles rest frame has reached a value sufficient to cause vacuum polarisation. \\
 
At SLAC-E144, a series of experiments were performed to test the nonlinearity of electron-photon and photon-photon interactions in strong electromagnetic fields \cite{Bamber:1999zt}. An intense laser beam was interacted with the high energy electron of the SLAC Final test beam. The nonlinearity parameters just introduced can be written as \cite{Bamber:1999zt, BulaE144Proposal}
\begin{equation}\label{EtaPar}
  \eta=e\frac{E_{\footnotesize\mbox{rms}\normalsize}}{\omega \,m }\,,\quad \Upsilon_e=\frac{E^{\ast}_{\small\mbox{rms}\normalsize}}{E_{\small\mbox{crit}\normalsize}} \,,\quad \Upsilon_{\gamma}=\frac{2\epsilon_{\gamma}}{m}\frac{E_{\small\mbox{rms}\normalsize}}{E_{\small\mbox{crit}\normalsize}} \,,
\end{equation}
where $\Upsilon_{\mbox{e}},\, \Upsilon_{\gamma} $ are nonlinearity parameters for incoming electrons and photons respectively, $E_{\footnotesize\mbox{rms}\normalsize}$ is the root-mean-square of the electric field of the laser, $\omega$ is its frequency, $E^{\ast}_{\footnotesize\mbox{rms}\normalsize}$ is the root-mean-square of the electric field in the electron rest frame and $\epsilon_{\gamma}$ is the energy of the photon. \\

These nonlinearity parameters depend on the intensity $I$ of the laser through $E_{\footnotesize\mbox{rms}\normalsize}=\sqrt{377[\Omega]\,I[\mbox{W/cm}^2]}$ \cite{McDonald:1986zz}: 
\begin{equation}
 \eta^2=3.7\cdot10^{-19}\,I\,\lambda^2\,, 
\end{equation}
where $I$ is written in W/cm$^2$ and $\lambda$, the wavelength of the laser, in $\mu$m. The intensity can be expressed in turn in terms of the energy of the laser beam $U$, the laser focus area $A$ and the pulse length $\tau$:
\begin{equation}\label{IntE144}
 I=\frac{U}{A\,\tau}\,\,\,\,\,\mbox{.}
\end{equation}

In Table \ref{ExpE144} we list the peak $\eta,\, \Upsilon_e,\,\Upsilon_{\gamma}$ measured at SLAC-E144 \cite{Bamber:1999zt}.

\begin{table}
\begin{center}\scriptsize
\begin{tabular}{l||c|c|c|c|c|c|}
&$\lambda$ (nm)&Measured peak $I$ (W/cm$^2$)&Measured peak $\mathbf{\eta}$&Measured peak $\mathbf{\Upsilon_e}$&Measured peak $\mathbf{\Upsilon_{\gamma}}$\\\hline\hline
\textbf{E144 green}&527&$\approx5\cdot10^{17}$&\cellcolor{blue!35}0.32&\cellcolor{red!35}0.27&\cellcolor{green!35}0.16\\\hline
\textbf{E144 IR}&1053  &$\approx5\cdot10^{17}$&\cellcolor{blue!35}0.4&\cellcolor{red!35}0.17&\cellcolor{green!35}0.08\\\hline
\end{tabular}   \normalsize     \end{center}
\caption{Peak nonlinearity parameters measured at SLAC-E144. }\label{ExpE144} 
\end{table}

\subsection{Nonlinear Compton scattering: $e^-+n\omega\to e^-+ \gamma$}

\begin{figure}[htb]\centering
\includegraphics[height=0.25\textwidth]{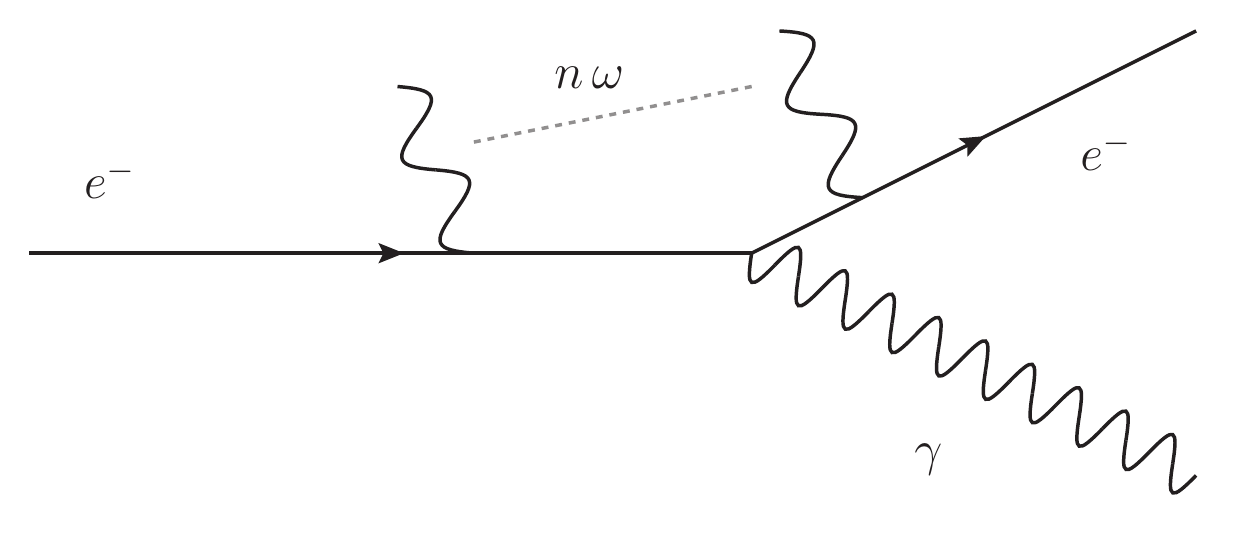}
\caption{Nonlinear Compton scattering.}\label{NLCSfeynman}
\end{figure}

The first nonlinear process studied at SLAC-E144 was the electron-photon interaction of nonlinear Compton scattering: $e^-+n\omega\to e^-+ \gamma$, represented in the Feynman diagram in figure \ref{NLCSfeynman}, where $p$ and $\mathcal{E}$ , $p^{\prime}$ and $\mathcal{E}^{\prime}$, $k$ and $\omega$, and $k^{\prime}$ and $\omega^{\prime}$ are the momenta and energies for the beam electron, scattered electron, laser photons and radiated photon, respectively.

\begin{figure}[htb]\centering
\includegraphics[height=0.25\textwidth]{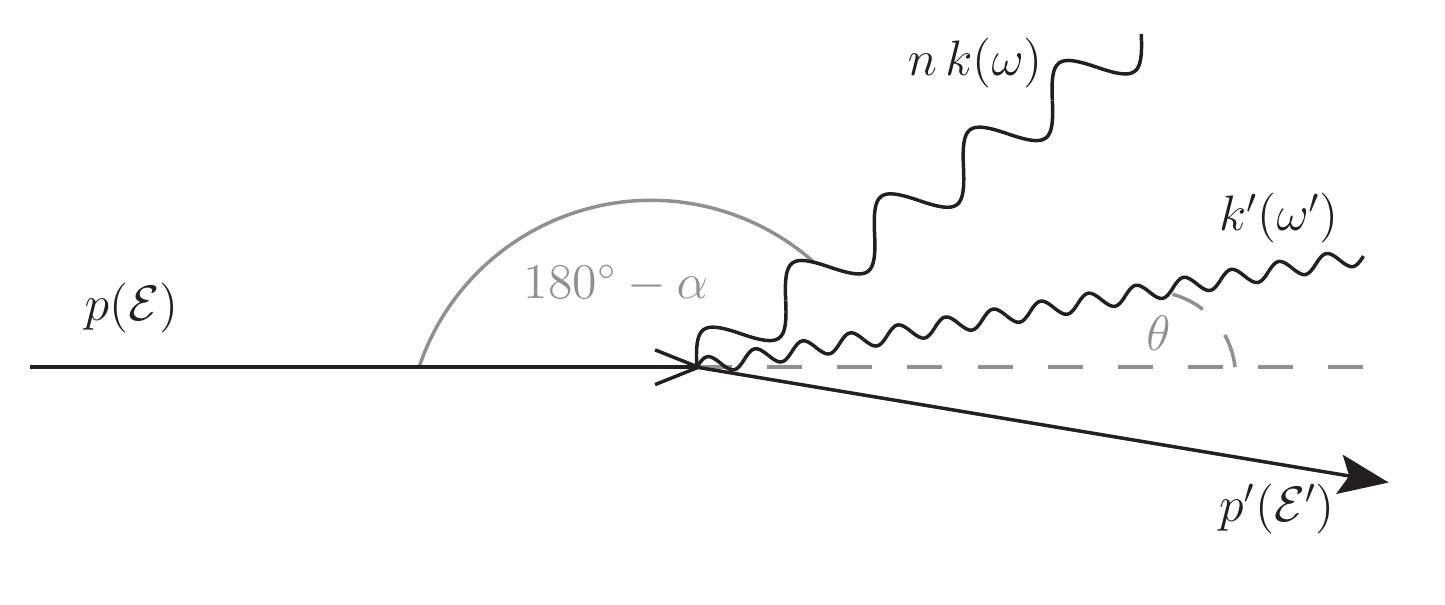}
\caption{Nonlinear Compton scattering dynamics.}\label{NLCSdynamics}
\end{figure}

In order to describe scattering processes in an external field environment \cite{Nikishov:1964zza,Nikishov:1964zz,Narozhnyi1964,Nikishov1965}, one has to apply the Furry Picture formalism \cite{Furry:1951zz}, which makes use of the electron wavefunctions that are solutions to the Dirac-Volkov equation of a charge in an external electromagnetic field \cite{Volkov35}. \\

Using the conventions in \cite{Bamber:1999zt}, we give, as an example, the differential rate for Compton scattering of an unpolarized beam electron by $n$ circularly polarized laser photons:


\begin{equation}
 \frac{dN_n(\omega^{\prime})}{d\omega^{\prime}}=\frac{\pi\,e^4 \,\rho_e\rho_{\omega}}{m^2\,\mathcal{E}^2 \,\omega}\left[-\frac{4}{\eta^2}J^2_n(z)+\left(2+\frac{u^2}{1+u}\right)[J_{n-1}^2(z)+J^2_{n+1}(z)-2J^2_{n}(z)]\right]\label{RateNLCS}\,,
\end{equation}

where $J_n(z)$ are first-type Bessel functions, $\rho_e$ and $\rho_{\omega}$ are the number density of beam electrons and of laser photons respectively, and

\begin{align}
u&=\frac{(k\cdot k^{\prime})}{(k\cdot p^{\prime})}\simeq\frac{\omega^{\prime}}{\mathcal{E}^{\prime}}\,,\\
 z&=\frac{2\eta}{u_1}\sqrt{\frac{u(u_n-u)}{1+\eta^2}}\,,\\
u_n&=n\,u_1\,,\\
 u_1&=\frac{2(k\cdot p)}{m^2(1+\eta^2)}\simeq \frac{2\omega\, \mathcal{E}(1+\beta\cos\alpha) }{m^2(1+\eta^2)}\,.
 \end{align}

\begin{figure}[htb]\centering
\includegraphics[height=0.25\textwidth]{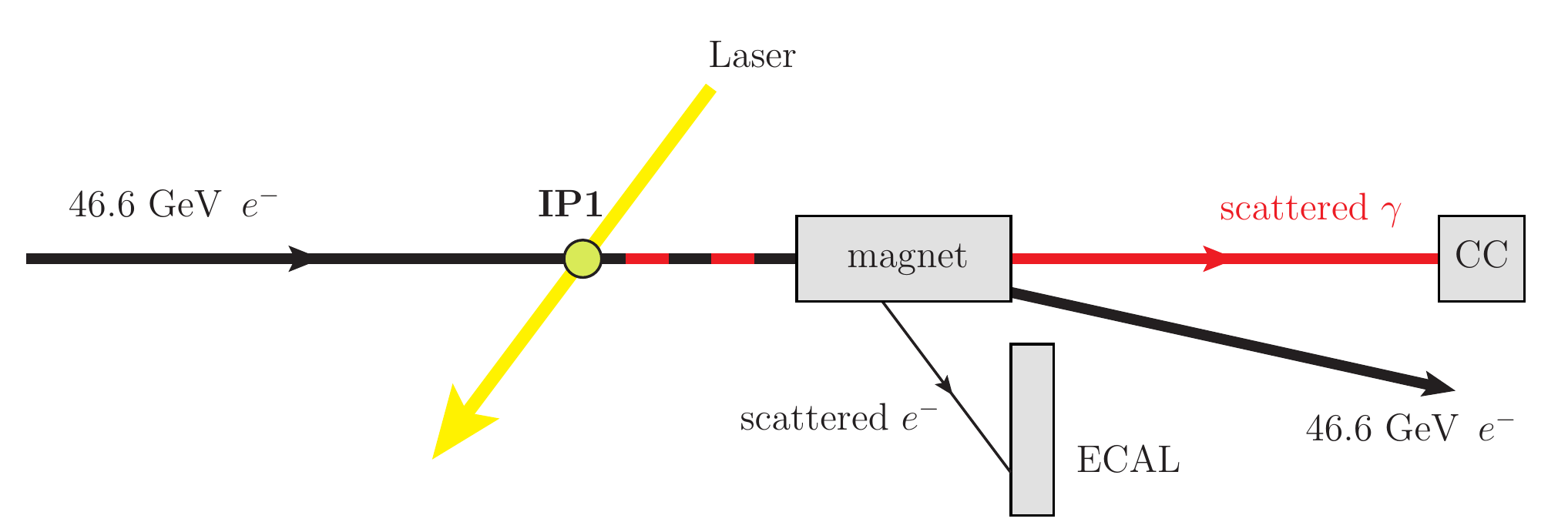}
\caption{Nonlinear Compton scattering setup.}\label{NLCS}
\end{figure}

Nonlinear Compton scattering, in which an electron is scattered by an external field from which it adsorbs $n$ photons emitting a final photon, was achieved at SLAC-E144 with the experimental setup depicted in Figure \ref{NLCS}.

The SLAC 46.6 GeV $e^-$ Final Test beam was collided, at an angle $\alpha=17^{\circ}$, with a laser beam operated at different frequencies ($\lambda_{green}$=527 nm $\lambda_{IR}$=1053 nm), energies (between 10 and 800 mJ), and alternatively with circular and linear polarizations at the interaction point IP1. The high energy Compton photons were back scattered along the direction of the incoming beam and were detected at a \u{C}erenkov counter (CC) after a magnet steered the residual 46.6 GeV beam electrons to a dump. Scattered Compton electrons were detected in an electron calorimeter (ECAL). \\
 
The main background process for nonlinear Compton scattering is Multiple Compton scattering, $e^-+n\omega\rightarrow e^-+m\omega^{\prime}$ which is energetically distinct and could therefore be clearly distinguished. The observed yield and spectra of scattered electrons at different laser intensities showed the occurrence of nonlinear Compton scattering, with an acceptable fit with the numerical simulations up to $n=3$ absorbed laser photons \cite{PhysRevLett.76.3116,Bamber:1999zt}. \\
 
There has recently been developments for the prospects of producing  Nonlinear Compton scatteringdue to improvments in ultrashort laser pulse technology \cite{Mackenroth:2010jr}.

\subsection{Breit-Wheeler pair production: $\gamma+n\omega\to e^{+}e^{-}$}

\begin{figure}[htb]\centering
\includegraphics[height=0.25\textwidth]{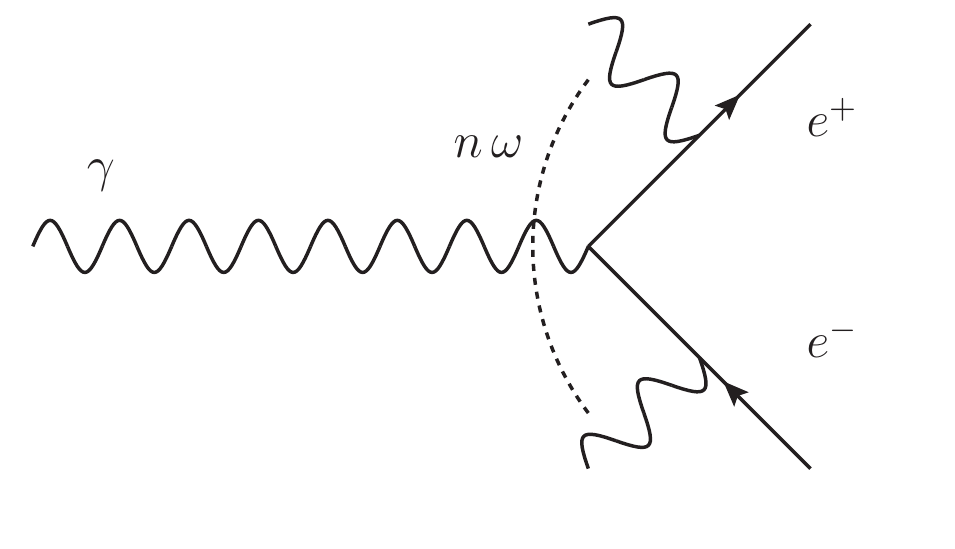}
\caption{Breit-Wheeler pair production according to the Furry Picture.}\label{BWPPfeynman}
\end{figure}

The nonlinear interaction of a photon with $n$ photons from an external electromagnetic field to produce an electron-positron pair was firstly addressed by \cite{Reiss1962}. The process $\gamma+n\omega\to e^{+}e^{-}$, represented by the Feynman diagram in Figure \ref{BWPPfeynman},

is related by a crossing symmetry to nonlinear Compton scattering. This symmetry results in a differential transition rate \eqref{RateBWPP} for the case of an unpolarized photon beam colliding with a circularly polarized laser \cite{Bamber:1999zt}, whose structure is similar to that of equation \eqref{RateNLCS}:

\begin{equation}
 \frac{dN_n(\mathcal{E}_{\pm})}{d\mathcal{E}_{\pm}}=\frac{2\pi\,e^4\,\rho_{\omega}\rho_{\omega^{\prime}}}{\omega\omega^{\prime\,2}}\left[\frac{2}{\eta^2}J_n(\zeta)+(2w-1)[J_{n-1}^2(\zeta)+J^2_{n+1}(\zeta)-2J^2_{n}(\zeta)]\right]\,,\label{RateBWPP}
\end{equation}

where

\begin{align}
w&=\frac{(k\cdot k^{\prime})^2}{4(k\cdot p_{-})(k\cdot p_{+})}\simeq\frac{\omega^{\prime\,2}}{4\mathcal{E}_{\pm}(\omega^{\prime}-\mathcal{E}_{\pm})}\,,\\
 \zeta&=\frac{2\eta}{w_1}\sqrt{\frac{w(w_n-w)}{1+\eta^2}}\,,\\
w_n&=n\,w_1\,,\\
w_1&=\frac{(k\cdot k^{\prime})}{2m^2(1+\eta^2)}\simeq \frac{\omega \omega^{\prime}(1+\beta\cos\alpha^{\prime}) }{2m^2(1+\eta^2)}\,\,.
 \end{align}
 
\begin{figure}[htb]\centering
\includegraphics[height=0.25\textwidth]{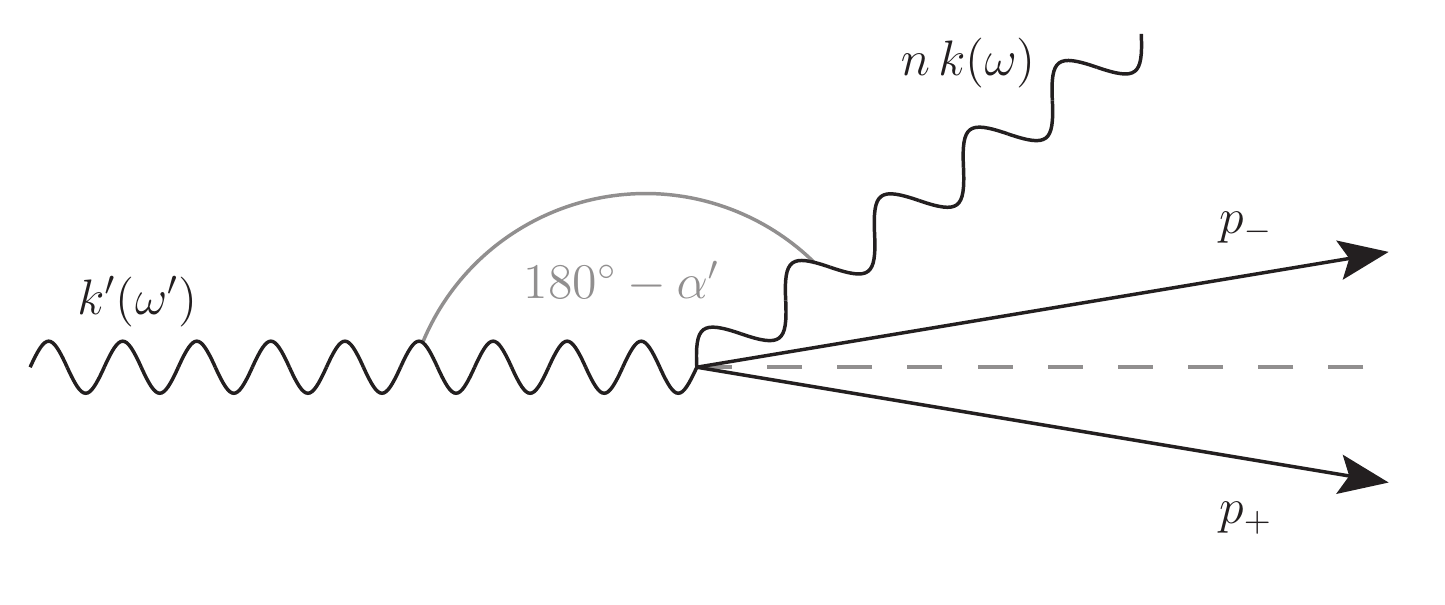}
\caption{Breit-Wheeler pair production dynamics.}\label{BWPPdynamics}
\end{figure}

Figure \ref{BWPPdynamics} represents Breit-Wheeler pair production where $\mathcal{E}_{\pm}, p_-,p_+$ are the energy and the momenta of the pair electron and positron, while $\omega,\,k$ and $\omega^{\prime},\,k^{\prime}$ are the momenta and energies of the laser photons and high energy beam photon respectively; $\rho_{\omega^{\prime}}$ is number density of beam photons.
 
\begin{figure}[htb]\centering
\includegraphics[height=0.25\textwidth]{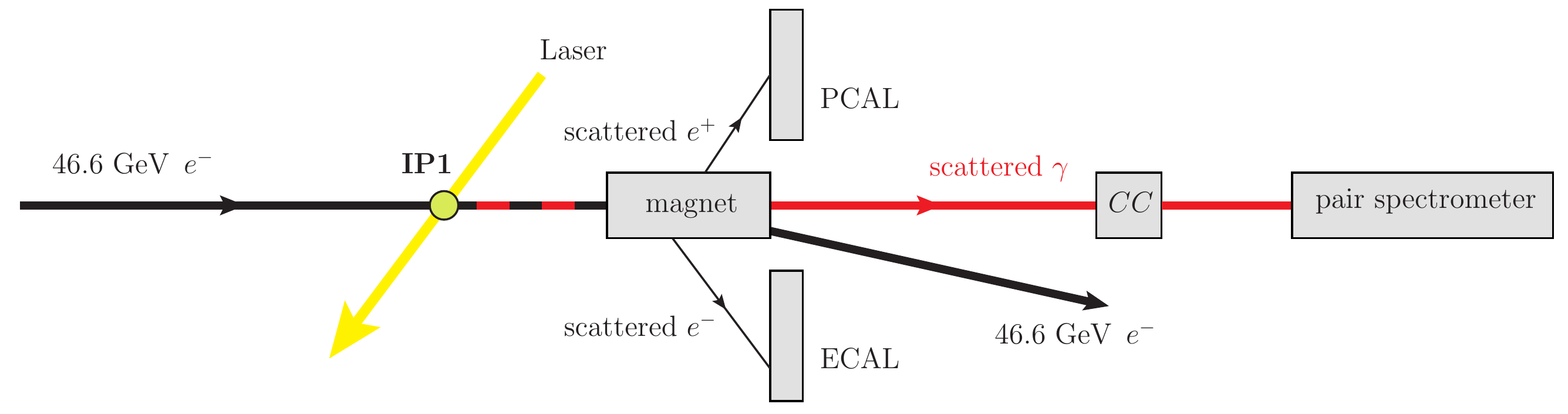}
\caption{Breit-Wheeler pair production setup.}\label{BWPP}
\end{figure}

The design of the Breit-Wheeler pair production experiment at SLAC-E144 is depicted in Figure \ref{BWPP}.

A dump magnet steered the residual 46.6 GeV electrons away from the beam line, while the scattered electrons and positrons were sent respectively to an electron (ECAL) and a positron (PCAL) calorimeter. The scattered photons were detected in a \u{C}erenkov counter (CC) or in pair spectrometer after being converted into electron-positron pairs. \\

The background trident process $e^-+n\omega\rightarrow e^-+ e^+e^-$ proved to be responsible for $<\!1$\% of detected positrons. Positron backgrounds from bremsstrahlung and Bethe-Heitler pair production were also considered. The observed spectra and yields of the scattered positrons unambiguously showed that at least four laser photons contributed to the production of $e^+e^-$-pairs \cite{PhysRevLett.79.1626,Bamber:1999zt}.

\section{Possibility at the ILC/CLIC}\label{SecILC}

We propose to set up a lepton-laser collision experiment at future linear collider, which builds on the success of SLAC-E144. This would enable us to explore strong fields physics with nonlinearity parameters exceeding those at SLAC-E144 by up to one order of magnitude. This would be possible because of the higher energy of the electron positron beams at the future linear collider as well as more intense, commercially available, lasers. \\

In Table \ref{ExpComparison} we compare the values of key parameters obtained at SLAC-E144 with those we expect (using relations \eqref{EtaPar}-\eqref{IntE144}) for the future linear collider based on International Linear Collider parameters \cite{Behnke:2013lya}. With the high energy beams of the ILC we would expect to exceed $E_{\footnotesize\mbox{cr}\normalsize}$ in the rest frame of the scattering electron even with the same IR laser of SLAC-E144. With off-the-shelf lasers available today, such as the PowerLite 9000 \cite{Powerlite}, and more modern laser optics for CPA and focussing, the strong field physics reach of the experiment can be further extended. \\

One of the interesting possibilities of considering such an experiment in the extraction line of the future linear collider is strong field experiements with positron initial states. The availability of positron bunches such the ones at the ILC would allow, to our knowledge, the first such of experiment conducted on positrons, resulting in a substantial improvement and innovation compared with the experiment described in \cite{Bulanov:2013cga}. Nonlinear effects from positron would add to tests of theoretical predictions of intense field, nonlinear QED. \\

There are several possibilities that can be considered for the location and realization of the proposed experiments at the ILC \cite{Adolphsen:2013kya,Behnke:2013lya} (or CLIC):

\begin{itemize}
 \item Locating the experiment in the polarimetry chicane of the extraction line, either making use of the polarimeter laser or a dedicated high intensity laser.

\item Using a high intensity laser in an upstream location, for example in the storage rings, and running it parasitically to the ILC operation.

\item As was the case for SLAC-E144, performing the experiments on an ILC test-beam.
\end{itemize}

\begin{table}
 \begin{center}\scriptsize
\begin{tabular}{l||c|c|c|c|c|c|c|c|c|}
&$\lambda$ (nm)&$\mathcal{E}_{\tiny\mbox{laser}\tiny}$ (J)&Focus &pulse (ps)&$I_{\tiny\mbox{peak}\scriptsize}$ (W/cm$^2$)&$E_{e^-}$ (GeV)&$\mathbf{\eta}$&$\mathbf{\Upsilon_e}$&$\mathbf{\Upsilon_{\gamma}}$\\\hline\hline
\textbf{E144 green}&527& 0.016\,-\,0.5&30 $\mu$m$^2$&1.5\,-\,2.5 &$\approx5\cdot10^{17}$&46.6&\cellcolor{blue!35}0.32&\cellcolor{red!35}0.27&\cellcolor{green!35}0.16\\\hline
\textbf{E144 IR}&1053 & 0.016\,-\,0.8 &60 $\mu$m$^2$&1.5\,-\,2.5 &$\approx5\cdot10^{17}$&46.6&\cellcolor{blue!35}0.4&\cellcolor{red!35}0.17&\cellcolor{green!35}0.08\\\hline
\textbf{ILC (E144 las.)}&1053 &0.8 &60 $\mu$m$^2$&1.5 &$9\cdot10^{17}$&125\,-\,500&\cellcolor{blue!35}0.60&\cellcolor{red!35}0.66\,-\,2.65&\cellcolor{green!35}0.46\,-\,2.38\\\hline
\textbf{ILC (PL$_{\mathbf{9000}}$)}&1064 &3 &40 $\mu$m$^2$&0.5 &$1.5\cdot10^{19}$&125\,-\,500&\cellcolor{blue!35}2.46&\cellcolor{red!35}2.72\,-\,10.88&\cellcolor{green!35}1.88\,-\,9.78\\\hline
\end{tabular}   \normalsize  \end{center}
\caption{Comparison between beams and peak nonlinearity parameters at SLAC-E144 and proposed tests at the ILC. }   \label{ExpComparison} 
\end{table}
\normalsize

A future specific experimental proposal will have to take into consideration effects such as electromagnetic cascades \cite{Bulanov:2013cga} or radiation interactions, see for example \cite{Neitz:2014hla}. We believe that the effect of the background process considered at SLAC-E144, like multiple Compton scattering (for recent theory developments see for example \cite{Mackenroth:2012rb}) and the trident process (see \cite{Hu:2010ye,Ilderton:2010wr,King:2013osa}), could be lowered thanks to higher statistics due to potentially longer running times.

\section{Mass-shift effects}\label{SecMassShift}
An electromagnetic charge in an external electromagnetic wave oscillates with the frequency of the external field. Since the amplitude of the oscillation is always smaller than the wavelength of the external wave \cite{McDonald:1986zz}, the oscillatory motion cannot be resolved by the light scattered by the quivering electron. The electron quiver motion appears as an angular and intensity dependent mass ``quasi-momentum'' $q$, and an attendant shift in its mass \cite{Brown:1964zzb,Kibble:1965zza}, 

\begin{gather}
 \overline{m}=m\sqrt{1+\eta^2},\,\,\,\,\,\,\,\,\,\,\eta=\frac{eE}{m\omega_0c}, \notag\\
 q_{\mu}=p_{\mu}+\frac{\eta^2m^2}{2k\cdot p}k_{\mu}, \quad q^2=\overline{m}^2\,.
\end{gather}

This effect has been explained not only classically but also by the solutions of Volkov-Dirac equations in a plane wave \cite{Volkov35}.\\

The dependence of the 1 vertex Compton scattering transition probability on the mass shift is obtained by a calculation within the Furry picture (equation \ref{RateNLCS}). The transition probability specifies a dependence on the mass shift exists in both the kinematics of the scattered electron and radiated photon. \\

The angular distribution and energy of a scattered electron was considered in an experiment at Rochester involving the ionization of Neon gas under the impact of an ultra intense laser \cite{Meyerhofer:96}. The electrons resulting from the interactions were studied in a spectrometer and the authors obtained a positive result for the mass shift effect. It has been shown that mass-shift effects in strong lasers can be controlled by pulse shaping and feasibly detected with current laser technology, as proposed by \cite{PhysRevLett.109.100402}. \\

An alternative experiment carried out by SLAC-E144, looked for a mass shift-dependent displacement of the Compton kinematic edge by analysing the energy distribution of the radiated photon. They were however unable to distinguish or rule out a mass-shift due to experimental limitations. We intend to repeat a similar experiment at a future linear collider with greater laser intensity and data collection statistics in order to settle the question.

\begin{figure}[htb]\centering
\includegraphics[height=0.4\textwidth]{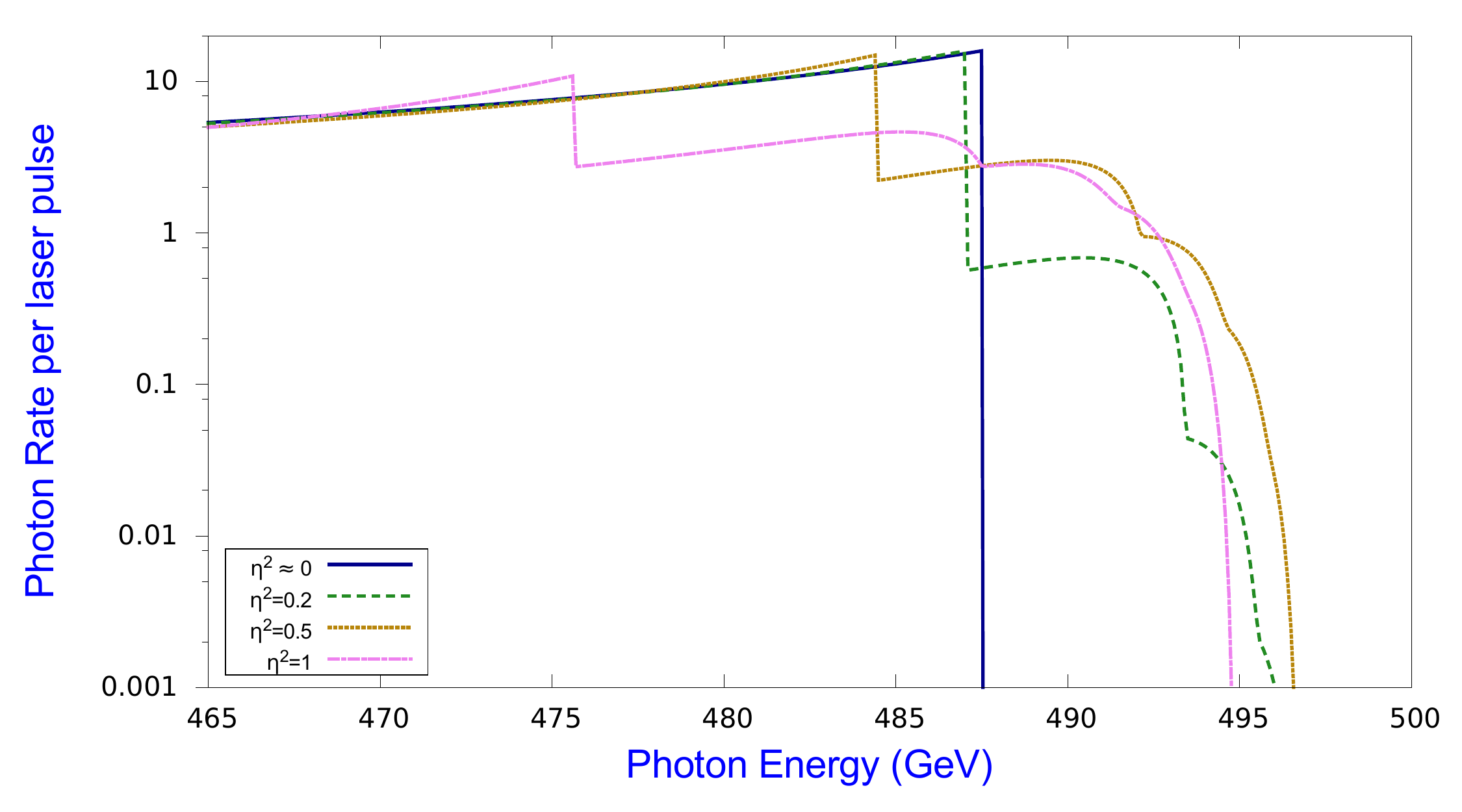}
\caption{Changes in the nonlinear Compton scattering kinematic edge due to intensity dependent mass shift and multiphoton effects.}\label{HICSsteprate}
\end{figure}

Figure \ref{HICSsteprate} shows the expected energy distribution of radiated photons for different laser intensities as expressed by the $\eta$ parameter. Here we assumed a 1 micron wavelength laser impinged on a 500 GeV electron bunch at a future linear collider with $10^7$ events per laser pulse. The plot shows a ``stepped'' multiphoton contribution with the mass shift effect seen in a horizontal shift of the leading Compton edge. \\

A positive result for the mass shift in a SLAC-E144 type experiment would confirm the result obtained at Rochester and would provide important confirmation of strong-field theory within the Furry picture.

\section{Strong laser/electron spin dependent effects}\label{SectionSpin}

The possibility of intense-laser/electron beam interactions in the extraction line of a future linear collider includes the possibility of spin dependent, strong field processes. To maximise the physics potential of a future collider, the beams will be polarized, including the possibility to flip the helicity and to depolarize the beams. This will allow a variety of spin states in the initial state of an electron beam/intense laser interaction. \\

There are two experimental effects to be discussed here. Firstly there is the quantum spin flip process which occurs at both the collider IP and an intense laser IP. In the first case the strong field is provided by the oncoming beam whose electromagnetic field can be described by a constant crossed plane wave. By contrast the laser IP will provide a strong circularly or linearly polarized field. \\

The spin dependent photon emission by an electron in the presence of a strong constant crossed plane wave field has been considered in the Furry picture formalism by \cite{Ritus:1972ky}. For an incident electron of momentum $p_\mu=(p_0,\overrightarrow{p})$ polarized in the $\overrightarrow{\xi}$ direction (in its rest frame), its spin 4-vector in a general frame is 

\begin{equation}\label{eq:Volkov}
s_\nu(\overrightarrow{\xi})=\left( \frac{\overrightarrow{\xi}\cdot \overrightarrow{p}}{m},\,\overrightarrow{\xi}+\frac{\overrightarrow{\xi}\cdot \overrightarrow{p}}{m(p_0+m)}\overrightarrow{p} \right)\,\,\,.
\end{equation}

The probability of photon emission is dependent on the spin vector, the strength of field and the relative propagation direction of the electron and the laser,

\begin{equation}
W(\overrightarrow{\xi})=-\frac{\alpha m^2}{\pi p_0}\int^\infty_0 \frac{d\upsilon}{(1+\upsilon)^2}\left[ \int dy + \frac{1+(1+\upsilon)^2}{y(1+\upsilon)}\frac{d}{dy} - \frac{yeF^*_{\mu\nu}p_\mu s_\nu}{m^3(1+\upsilon)}\right] \mbox{Ai}(y) \, ,
\end{equation}
where
\begin{equation}
 \quad y=\frac{u\, m^3}{\sqrt{(eF_{\mu\nu} p_\nu)^2}}\,\,\,. \notag
\end{equation}

The probability of photon emission contains integrals and derivatives of the analytic function related to the symmetry of the external field, namely the Airy function corresponding to a constant crossed plane wave. For a laser field, the same analysis can be performed. Here the Bessel function of the first kind will appear in the stead of the Airy function. \\

Detection of spin dependent effects in this process at a laser/electron IP will have a direct bearing on the collider IP effects to the extent that the numerical values of the Bessel function can be compared to that of the Airy function. This methodology can serve primarily as a confirmation of the Furry picture formalism and as a cross check to collider IP effects measured by different means.

\section{Conclusions}\label{SecConclusions}
The future linear collider, apart from physics events at its primary interaction point, offers the opportunity to perform a set of strong field experiments using an intense laser in the extraction line of the electron and/or positron beam. Such an experiment would reach even further into the strong field regime than previous successful experiments such as those at SLAC-E144. New strong field effects such as those from spin dependent processes could be observed as well settling the long standing question of the mass shift. Such experiments would also shed light on the strong field beam-beam effects at the primary interaction point. \\

A dedicated experimental proposal which would detail the experimental setup, and simulate transition rates together with and possible backgrounds would be worth considering in the near future.

\section*{Acknowledgements}

S.~P. has been supported by DFG through the grant SFB 676 ``Particles, Strings, and the Early Universe''.
    
\bibliography{LCWS_E144}
\bibliographystyle{JHEP}

\end{document}